\documentclass{moriond}

\usepackage[utf8]{inputenc} 
\usepackage{amsmath}

\def\Journal#1#2#3#4{{#1} {\bf #2}, #3 (#4)}

\def\NPB{{\em Nucl. Phys.} B}

\def\PRL{\em Phys. Rev. Lett.}
\def\RMP{\em Rev. Mod. Phys.}

\def\be{\begin{equation}}
\def\ee{\end{equation}}
\def\bea{\begin{eqnarray}}
\def\eea{\end{eqnarray}}

\newcommand{\Photo}{}

\begin{document}
\vspace*{4cm}
\title{ELECTRIC DIPOLE MOMENTS AND NEUTRINO MASS MODELS}

\author{ SELIM TOUATI }

\address{Laboratoire de Physique Subatomique et de Cosmologie,
Université Grenoble-Alpes, CNRS/IN2P3,\\53 avenue des Martyrs,\\ 
38026 Grenoble Cedex, France}

\maketitle\abstracts{
While Jarlskog-like flavor invariants are adequate for estimating CP-violation from closed fermion loops, non-invariant structures arise from rainbow-like processes.
For the CKM contributions to the quark electric dipole moments (EDMs), or the PMNS contributions to lepton EDMs, the dominant diagrams have a rainbow topology whose flavor structure does not collapse to flavor invariants. Numerically, they are found typically much larger, and not necessarily correlated with, Jarlskog-like invariants.
The flavor structures in the quark and lepton sectors are systematically studied, assuming different mechanisms for generating neutrino masses.
In addition, the combined study of both Jarlskog-like and rainbow-like flavor structures shed new lights on the possible correlations between quark and lepton EDMs. \cite{SmithTouati}}

\section{EDMs generated by the CKM phase}

In the standard model (SM), the only source of weak CP-violation is the complex phase of the CKM matrix. In order to measure the strength of CP-violation, one can construct a flavor invariant (basis-independent) which is sensitive to this phase, called the Jarlskog invariant \cite{Jarl}. A non-vanishing Jarlskog invariant is a necessary condition for having CP-violation. In the SM, all CP-violating effects are proportional to this invariant.
However, this invariant is adequate for estimating CP-violation from closed fermion loops. For example, let us consider the CKM-induced lepton EDMs. Because the leptons cannot feel directly the complex phase of the CKM matrix, we need to go through a closed quark loop. The dominant diagram is:

\begin{figure}[h!]
\centering
\includegraphics[scale=0.6]{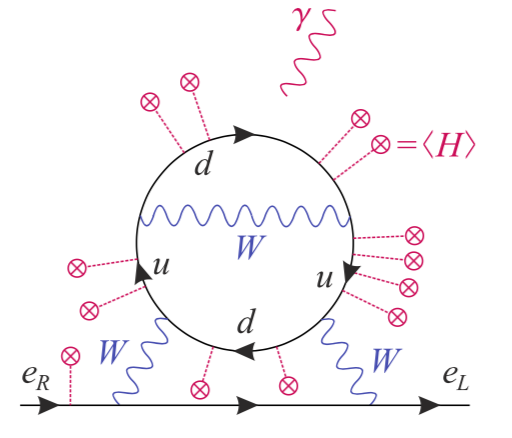}
\caption{CKM-induced lepton EDM}
\label{fig:CKMleptonEDM}
\end{figure}
This EDM is tuned by the Jarlskog invariant $\det[Y_{u}^{\dagger}Y_{u},Y_{d}^{\dagger}Y_{d}]$ which is proportional to the imaginary part of a quartet $Im(V_{us}V_{cb}V_{ub}^{\ast}V_{cs}^{\ast})$. As for the quarks, they can feel directly the complex phase of the CKM matrix and then there are non-invariants structures which arise from rainbow-like processes. Indeed, the dominant diagrams for the CKM-induced quark EDMs have a rainbow topology. For instance, for the d-quark EDM:

\begin{figure}[h!]
\centering
\includegraphics[scale=0.6]{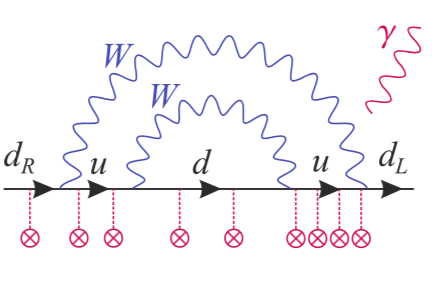}
\caption{CKM-induced d-quark EDM}
\label{fig:CKMquarkEDM}
\end{figure}

This EDM is tuned by the imaginary part of the 1-1 entry of a non-invariant commutator $Im(\textbf{X}^{dd}_{q})$, where:
\begin{equation}
\textbf{X}_{q}=\mathbf{[Y}_{u}^{\dagger}\mathbf{Y}_{u},\mathbf{Y}_{u}^{\dagger}\mathbf{Y}_{u}\mathbf{Y}_{d}^{\dagger}\mathbf{Y}_{d}\mathbf{Y}_{u}^{\dagger}\mathbf{Y}_{u}],
\label{eq:Xq}
\end{equation} which is also proportional to $Im(V_{us}V_{cb}V_{ub}^{\ast}V_{cs}^{\ast})$ as for the lepton EDMs (because we are in the SM), but not with the same proportionality factor. It turns out to be much larger by 10 orders of magnitude: 
\begin{equation}
Im\mathbf{[Y}_{u}^{\dagger}\mathbf{Y}_{u},\mathbf{Y}_{u}^{\dagger}\mathbf{Y}_{u}\mathbf{Y}_{d}^{\dagger}\mathbf{Y}_{d}\mathbf{Y}_{u}^{\dagger}\mathbf{Y}_{u}]^{dd}\gg\det[Y_{u}^{\dagger}Y_{u},Y_{d}^{\dagger}Y_{d}].
\end{equation} In the SM, the rainbow-like flavor structures are typically much larger than the invariant determinants and they are correlated (strictly proportional). 
Now, let us turn on neutrino masses (beyond the SM) and check whether this behavior is confirmed or not. As we do not know yet the nature of the neutrino (Dirac or Majorana particle), we will consider both scenarios for generating neutrino masses.

\section{EDMs in the presence of neutrino masses}

\subsection{Dirac neutrino masses}

\begin{figure}[h!]
\centering
\includegraphics[scale=0.6]{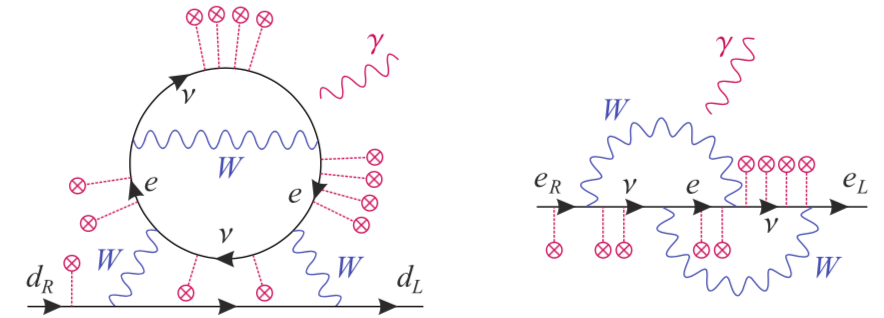}
\caption{PMNS-induced quark (on the left) and lepton (on the right) EDMs}
\label{fig:DiracEDMs}
\end{figure}

The simplest way of including neutrino masses to the SM is to extend its particles content by adding three right-handed (RH) fully neutral neutrinos (one for each generation). They belong to the trivial representation of the SM gauge group: $N=\nu_{R}^{\dagger}\sim(1,1)_{0}$. 
We add to the SM Yukawa Lagrangian an extra Yukawa interaction for the neutrinos:

\begin{equation}
\mathcal{L}_{Yukawa}=\mathcal{L}_{Yukawa}^{SM}-N^{I}Y_{\nu}^{IJ}L^{J}H^{\dagger C}+h.c.
\end{equation} We have a new neutrinos-related flavor structure $Y_{\nu}$ ($3\times3$ matrix in flavor space). In the presence of neutrino masses, we get an additional source of weak CP-violation coming from the complex phase of the PMNS matrix. In complete analogy with the quark sector, we can construct new CP-violating flavor structures which tune the PMNS-induced quark and lepton EDMs. In this case, quark EDMs have a bubble topology whereas lepton EDMs have a rainbow topology. For instance, the dominant diagrams for the PMNS-induced quark and lepton EDMs are shown in figure \ref{fig:DiracEDMs}.
They are tuned respectively by $J_{\mathcal{CP}}^{Dirac}$ and $Im(\textbf{X}_{e}^{Dirac})^{11}$, where 
\begin{align}
J_{\mathcal{CP}}^{Dirac}= & \frac{1}{2i}\det\left[Y_{\nu}^{\dagger}Y_{\nu},Y_{e}^{\dagger}Y_{e}\right]\\
\textbf{X}_{e}^{Dirac}= & \left[Y_{\nu}^{\dagger}Y_{\nu},Y_{\nu}^{\dagger}Y_{\nu}Y_{e}^{\dagger}Y_{e}Y_{\nu}^{\dagger}Y_{\nu}\right].
\label{eq:XeDirac}
\end{align}
In this scenario, $Im(\textbf{X}_{e}^{Dirac})^{11}$ is 11 orders of magnitude larger than $J_{\mathcal{CP}}^{Dirac}$ and they are correlated (strictly proportional).

\subsection{Majorana neutrino masses}

\begin{figure}[h!]
\centering
\includegraphics[scale=0.7]{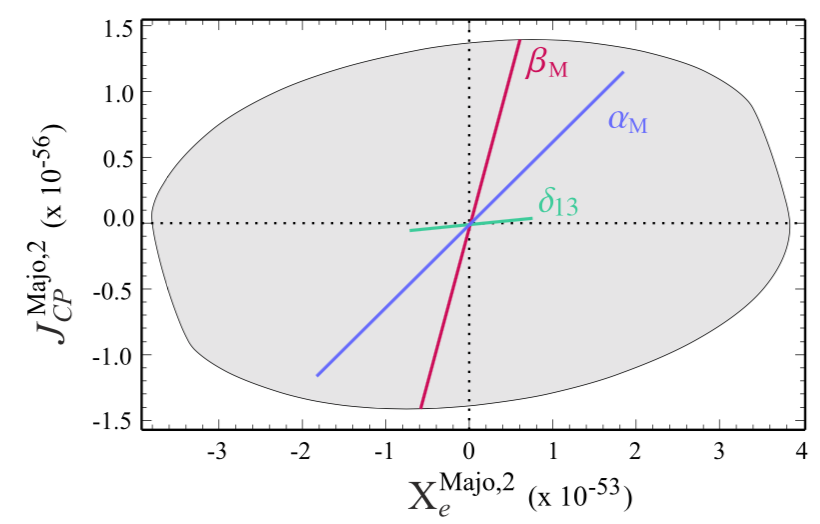}
\caption{Area spanned by $J_{\mathcal{CP}}^{Majo}$ and $Im(\textbf{X}_{e}^{Majo})^{11}$}
\label{fig:CorrelationMajo}
\end{figure}

Another way for generating neutrino masses is possible if we consider  Majorana masses. In this mechanism, there is no additional RH neutrinos, we get directly a gauge-invariant but lepton-number violating mass term for the left-handed (LH) neutrinos. Indeed, we add to the SM Yukawa Lagrangian the effective dimension-five Weinberg operator:

\begin{equation}
\mathcal{L}_{Yukawa}=\mathcal{L}_{Yukawa}^{SM}-\frac{1}{2v}(L^{I}H)(\Upsilon_{\nu})^{IJ}(L^{J}H)+h.c,
\end{equation} which after spontaneous symmetry breaking collapses to a Majorana mass term for the LH neutrinos:

\begin{equation}
\frac{1}{2v}(L^{I}H)(\Upsilon_{\nu})^{IJ}(L^{J}H)\overset{SSB}{\longrightarrow}\frac{v}{2}(\Upsilon_{\nu})^{IJ}\nu_{L}^{I}\nu_{L}^{J}.
\end{equation}
$\Upsilon_{\nu}$ (3$\times$3 matrix in flavor space) is a new flavor structure purely of the Majorana type. In this model, we must redefine the PMNS matrix in order to add two new CP-violating phases, called Majorana phases, 
\begin{equation}
U_{PMNS}\rightarrow U_{PMNS}\cdot diag(1,e^{i\alpha_{M}},e^{i\beta_{M}}).
\end{equation}
Let us consider the PMNS-induced quark and lepton EDMs in this scenario. The dominant diagrams are:

\begin{figure}[h!]
\centering
\includegraphics[scale=0.6]{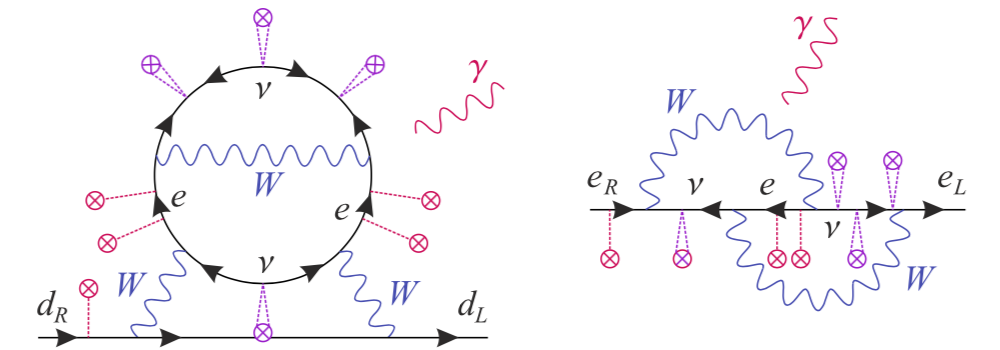}
\caption{PMNS-induced quark (on the left) and lepton (on the right) EDMs}
\label{fig:MajoEDMs}
\end{figure} 
The CP-violating flavor structures which tune these EDMs are $J_{\mathcal{CP}}^{\mathrm{Majo}}$  \cite{Branco} and $Im(\mathbf{X}_{e}^{\mathrm{Majo}})^{11}$, where:
\begin{align}
J_{\mathcal{CP}}^{\mathrm{Majo}}= & \frac{1}{2i}Tr[\mathbf{\Upsilon}_{\nu}^{\dagger}\mathbf{\Upsilon}_{\nu}\cdot\mathbf{Y}_{e}^{\dagger}\mathbf{Y}_{e}\cdot\mathbf{\Upsilon}_{\nu}^{\dagger}(\mathbf{Y}_{e}^{\dagger}\mathbf{Y}_{e})^{T}\mathbf{\Upsilon}_{\nu}-\mathbf{\Upsilon}_{\nu}^{\dagger}(\mathbf{Y}_{e}^{\dagger}\mathbf{Y}_{e})^{T}\mathbf{\Upsilon}_{\nu}\cdot\mathbf{Y}_{e}^{\dagger}\mathbf{Y}_{e}\cdot\mathbf{\Upsilon}_{\nu}^{\dagger}\mathbf{\Upsilon}_{\nu}]\\
\mathbf{X}_{e}^{\mathrm{Majo}}= & [\mathbf{\Upsilon}_{\nu}^{\dagger}\mathbf{\Upsilon}_{\nu},\mathbf{\Upsilon}_{\nu}^{\dagger}(\mathbf{Y}_{e}^{\dagger}\mathbf{Y}_{e})^{T}\mathbf{\Upsilon}_{\nu}].
\end{align} 
We find that $Im(\textbf{X}_{e}^{Majo})^{11}$ is 4 orders of magnitude larger than $J_{\mathcal{CP}}^{Majo}$ but in this scenario they are not correlated. In figure \ref{fig:CorrelationMajo}, we can see the values that can take the PMNS-induced quark and lepton EDMs (tuned respectively by $J_{\mathcal{CP}}^{Majo}$ and $Im(\textbf{X}_{e}^{Majo})^{11}$). The lightest neutrino mass $m_{\nu 1}$ is set to $1eV$ and the CP-violating phases (PMNS phase $\delta_{13}$ and the Majorana phases $\alpha_{M}$ and $\beta_{M}$) are allowed to take on any values.
The lines show the strict correlation occuring when only one phase is non-zero. When the three phases are into action, because the flavor structures have different dependences in these phases, the result is that the quark and lepton EDMs become decorrelated.

\section{Conclusion}

In the paper \cite{SmithTouati}, we developped a systematic method to study the flavor structure behind the quark and lepton EDMs which can be extended easily to other more complicated models (Sterile neutrinos, SUSY etc...).
The rainbow-like non-invariant flavor structures are found to be typically much larger than the Jarlskog-like flavor invariants. Interestingly, we find a different behavior for Dirac and Majorana neutrinos. Quark and lepton EDMs are proportional in the former case whereas they are completely independent in the latter case. Indeed, quark and lepton EDMs have different dependences on Majorana phases. Finally, by studying the flavor structures behind the quark and lepton EDMs, we get the relations shown in table \ref{tab:SumRules} between EDMs of different generations.
\begin{table}[h!]
\caption[]{Sum rules}
\label{tab:SumRules}
\vspace{0.4cm}
\begin{center}
\renewcommand{\arraystretch}{2.0}
\begin{tabular}{ c | c | c |}
\cline{2-3}
 & CKM-induced EDMs & PMNS-induced EDMs \\
\hline
\multicolumn{1}{|c|}{Quarks} & $\frac{d_{d}}{m_{d}}+\frac{d_{s}}{m_{s}}+\frac{d_{b}}{m_{b}}=0$ & $\frac{d_{d}}{m_{d}}=\frac{d_{s}}{m_{s}}=\frac{d_{b}}{m_{b}}$\\
\hline
\multicolumn{1}{|c|}{Leptons} & $\frac{d_{e}}{m_{e}}=\frac{d_{\mu}}{m_{\mu}}=\frac{d_{\tau}}{m_{\tau}}$ & $\frac{d_{e}}{m_{e}}+\frac{d_{\mu}}{m_{\mu}}+\frac{d_{\tau}}{m_{\tau}}=0$ \\
\hline
\end{tabular}
\end{center}
\end{table} For example, the CKM-induced quark EDMs and the PMNS-induced lepton EDMs are tuned by the non-invariant commutators \ref{eq:Xq} and \ref{eq:XeDirac} and because a commutator is traceless, we get these sum rules.

\end{document}